\documentclass[10pt,oneside,twocolumn,pra]{revtex4}

\usepackage{revsymb,graphicx}

\usepackage[english]{babel}

\begin{document}

\title{Weak Values and Continuous-Variable Entanglement Concentration}

\author{David Menzies and Natalia Korolkova}
\email{nvk@st-andrews.ac.uk} \affiliation{School of Physics and
Astronomy, University of St. Andrews, North Haugh, St. Andrews KY16
9SS, UK.}
\date{\today}

\begin{abstract}
We demonstrate a general weak measurement model which allows
Gaussian preserving entanglement concentration of the two mode
squeezed vacuum. The power of this simple and elegant protocol is
through the constraints it places on possible ancilla states and
measurement strategies that will allow entanglement concentration.
In particular, it is shown how previously discovered protocols of
this kind emerge as special examples of the general model described
here. Finally, as evidence of its utility, we use it to provide
another novel example of such a protocol.
\end{abstract}

\maketitle

\section{Introduction}

As any textbook on quantum mechanics will testify, the measurable
values of any observable ${\cal O}$  coincide with the eigenvalues
of an appropriate self-adjoint operator $\hat O$. However, in
\cite{aharonov&vaidman:1990}, {\it Aharonov} and {\it Vaidmann}
showed that if a system with observable ${\cal O}$ is both pre and
post-selected in the states $\vert \Phi_1 \rangle$ and $\vert \Phi_2
\rangle$, then the observable can possess so-called weak values:
\begin{equation}
O_W = \frac{\langle \Phi_1 \vert {\hat O} \vert \Phi_2
\rangle}{\langle \Phi_1 \vert \Phi_2 \rangle}. \label{eqn1}
\end{equation}
It is vital to note that (\ref{eqn1}) does not coincide
with the eigenvalues of the operator $\hat O$ and can in general be
complex. One can transfer the weak values belonging to a particular
system onto another via indirect quantum measurement as introduced
by {\it Von Neumann} \cite{vonneumman:1955}, with a caveat - the
coupling strength between signal and probe must be weak
\cite{aharonov&vaidman:1990}. Such models are collectively, rather
appropriately, known as {\it weak measurements}. Despite initial
controversy, the concept of weak values and measurements have
enjoyed a surge of theoretical development
\cite{aharonov&vaidman:1990,johansen:2004,aharonav:2005,story:1991,agarwal:2007},
and eventual experimental confirmation \cite{pryde:2005}.

Perhaps unexpectedly, weak measurements have an application in the
currently open problem of entanglement distillation of Gaussian
states. So far, one possible approach was provided by distillation
protocols involving the subtraction of a definite number of photons
\cite{browne:2003,eisert:2004}. In contrast, the model reported here
provides a general method for {\it Procrustean} entanglement
distillation protocols which subtract or add an {\it indefinite}
number of photons \cite{fiurasek:2003,menzies:2006}. These latter
protocols probabilistically modify the average number of photons of
a particular entangled Gaussian state whilst preserving its
essential Gaussian features. Distillation of entangled resources is
a major issue in quantum information processing, in particular for
the development of quantum repeaters that allow for fault tolerant
communication between different information processors. In this
paper we address these tasks using mesoscopic, continuous variables,
a viable and extremely promising alternative to the traditional
quantum bit-based information processing.

We have discovered that \cite{fiurasek:2003,menzies:2006} are
special cases of a general weak measurement interaction. Using the
weak value paradigm, we demonstrate how to construct a general model
of such {\it Procrustean} protocols. Moreover, we identify that the
features of these protocols, namely success conditions and Gaussian
preservation are not unique to the particular choices advocated in
both \cite{fiurasek:2003,menzies:2006}. Instead, our general
analysis reveals that the origin of these features lie with the
consequences of performing a weak measurement. Furthermore, our
model constrains the pre and post-selected ancilla states whilst
providing a method for determining which possible combinations work.
Indeed, \cite{fiurasek:2003,menzies:2006} emerge as examples of a
general rule and we provide another example which of a possible
configuration.

The protocols \cite{fiurasek:2003,menzies:2006} were suggested as
possible resolutions to the following quantum communication
scenario. Suppose Alice and Bob wish to perform a particular
continuous-variable entanglement assisted protocol. Further assume
that they share a two-mode squeezed vacuum (TMSS) encoded in two
light modes, as an entanglement resource
\cite{barnett&radmore:1997,gerry&knight:2005}:
\begin{equation}
\vert \zeta (\lambda) \rangle = \sqrt{1- \lambda^2}
\sum_{n=0}^{\infty} \lambda^n \vert n , n \rangle.\label{eqn2}
\end{equation}
To ensure maximum performance from their protocol, they
must possess a high quality entangled state. Ergo, they wish to
increase the entanglement of their shared entangled state before
consuming it. Moreover, they require this to be done in a manner
that will preserve its Gaussian features. Accordingly, they want to
probabilistically map their initial TMSS $\vert \zeta(\lambda)
\rangle$ to another more entangled one $\vert \zeta(\lambda')
\rangle$.

\section{The Protocol}

To solve this problem we use the {\it Procrustean} method
\cite{bennett:1995}, where we probabilistically modify the {\it
Schmidt} coefficients of the input state to generate an output state
with a greater degree of entanglement whilst preserving the basis.
The following configuration is advocated, as depicted in fig(1), the
entangled state in modes $A$ and $B$ is coupled to ancilla state in
mode $C$ by means of a unitary evolution between $B$ and $C$. The
requirements of the {\it Procrustean} method dictate that the
interaction Hamiltonian describing this process must be of the form
\begin{equation}
\hat H_I = \hbar \kappa(t) {\hat n}_B \otimes {\hat O}_C.
\label{eqn3}
\end{equation}
This is required to preserve the {\it Schmidt} basis of
the TMSS, i.e. the {\it Fock} basis. Assuming the interaction
persists for $T$ seconds, then the corresponding unitary evolution
operator is
\begin{equation}
\hat U = e^{- i \int_0^T \kappa(t) {\hat n}_B {\hat O}_c} = e^{-i
\kappa_T {\hat n}_B {\hat O}_C }, \label{eqn4}
\end{equation}
where $\kappa_T = \kappa(T)-\kappa(0)$.

Following this, Bob performs a measurement on the ancilla and
post-selects it in the state $\vert \Phi_2 \rangle$. Consequently,
the state shared between Alice and Bob is given by
\begin{eqnarray}
\vert \Psi_f \rangle = {\cal N} \langle \Phi_2 \vert e^{-i \kappa_T
{\hat n}_B {\hat O}_C } \vert \zeta(\lambda) \rangle \vert \Phi_1
\rangle \nonumber \\
= {\cal N'} \sum_{m=0}^{\infty} \frac{(- i \kappa_T)^m}{m!}
\frac{\langle \Phi_2 \vert {\hat O}_C^m \vert \Phi_1 \rangle}{
\langle \Phi_2 \vert \Phi_1 \rangle} {\hat n}_B^m \vert
\zeta(\lambda) \rangle. \label{eqn5}
\end{eqnarray}
The weak value of $\hat O_C$ is defined as
\begin{equation}
O_W = \frac{\langle \Phi_2 \vert \hat O_C \vert \Phi_1
\rangle}{\langle \Phi_2 \vert \Phi_1 \rangle}, \label{eqn6}
\end{equation}
and so the final state of the system is given as
\begin{equation}
\vert \Psi_f \rangle = {\cal N'} \exp \left ( - i \kappa_T O_W
\hat{n}_B \right )\vert \zeta(\lambda) \rangle, \label{eqn7}
\end{equation}
if the weakness condition
\begin{equation}
\sum_{m=2}^{\infty} \frac{(-i \kappa_T)^m}{m!} \left \{ O^m_W -
(O_W)^m \right \} \hat n_b^m \vert \zeta(\lambda) \rangle \approx 0
\vert \phi \rangle \label{eqn8}
\end{equation}
is obeyed. Here $\vert \phi \rangle$ is an arbitrary
vector in ${\cal H}_A \otimes {\cal H}_B$ and $O^m_W = \langle
\Phi_2 \vert \hat O_C^m \vert \Phi_1 \rangle / \langle \Phi_2 \vert
\Phi_1 \rangle$. By using the linear independence of the {\it
Schmidt} basis of the TMSS we can express (\ref{eqn8}) as set of
equations:
\begin{equation}
\lambda^n \left ( \frac{\langle \Phi_2 \vert e^{ - i \kappa_T n
{\hat O}_C} \vert \Phi_1 \rangle}{\langle \Phi_2 \vert \Phi_1
\rangle } - e^{- i \kappa_T n O_W} \right ) \approx 0 \; \; \forall
n \in [0, \infty). \label{eqn9}
\end{equation}

Assuming that the above weakness condition is satisfied means that
the output state is another TMSS as (\ref{eqn7}) yields
\begin{equation}
\vert \Psi_f \rangle = \sqrt{1 - \lambda^2 e^{2 \kappa_T Img(O_W)}}
\sum_{n=0}^{\infty} \lambda^n e^{- i \kappa_T O_W n} \vert n ,n
\rangle, \label{eqn10}
\end{equation}
This only holds subject to $\lambda^2 e^{2 Img(O_W) \kappa_T} < 1$,
otherwise the output state is un-physical as the normalisation
constant will not converge. From (\ref{eqn10}) it can be seen that
the real part $O_W$ induces a phase shift on the TMSS whereas the
imaginary part modifies the average number of photons in the state.
Put succinctly, the induced transformation is $\lambda \rightarrow
\lambda' = \lambda e^{- i \kappa_T O_W}$. Thus, the average number
of photons has been altered \cite{barnett&radmore:1997},
\begin{equation}
\frac{2 \lambda^2}{1 -\lambda^2} \rightarrow \frac{2 \lambda^2
e^{2\kappa_T \mbox{Img}(O_W)}}{1 -\lambda^2e^{2\kappa_T
\mbox{Img}(O_W)}}, \label{eqn11}
\end{equation}
and as a result the entanglement content of the state is
also modified. It is in this sense that we can subtract or add an
indefinite number of photons to our target state.
\begin{figure}[b]
\centering
\includegraphics[height=40mm]{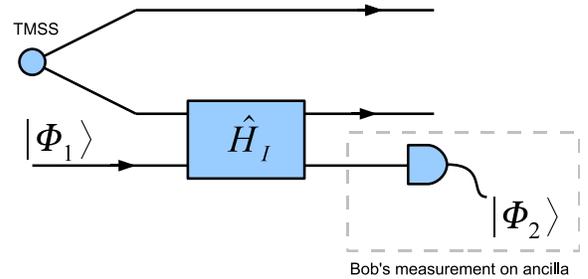}
\caption{\it Bob mixes his half of the TMSS with an ancillary mode
pre-selected in $\vert \Phi_1 \rangle$ via an non-linear photonic
interaction described by the Hamiltonian ${\hat H}_I$. The ancilla
mode is then subjected to a post-selected measurement leaving it in
the state $\vert \Phi_2 \rangle$. }
\end{figure}
To verify entanglement concentration, we use method of majorization.
Let ${\bf d} = (d_0^2, d_1^2, \ldots )^T$ be the ordered vector of
the eigenvalues of the reduced density matrices of (\ref{eqn7}) and
${\bf c}  = (c_0^2, c_1^2, \ldots )^T$ be the analogues object for
the initial TMSS. Then (\ref{eqn7}) is more entangled than $\vert
\zeta(\lambda) \rangle$ if its reduced density matrices are more
mixed. This will be the case if ${\bf d}$ is {\it majorized} by
${\bf c}$, which is written as ${\bf d} \prec {\bf c}$ and defined
by \cite{hayashi:2006,zyczkowski:2002}
\begin{equation}
\sum_{k=\ell}^{\infty} d_k^2 \geq \sum_{k=\ell}^{\infty} c_k^2,
\label{eqn12}
\end{equation}
for $\ell \in [1,\infty)$.  This follows since measures of
bipartite pure state entanglement such as the {\it Von Neumann}
entropy and the purity belong to the {\it Shur concave}
\cite{zyczkowski:2002} and hence, preserve the majorization order
\begin{equation}
{\bf d} \prec {\bf c} \Longrightarrow f({\bf d}) \geq f({\bf c}).
\label{eqn13}
\end{equation}
It is sufficient for entanglement concentration to show
that the eigenvalues of the reduced density matrices of the output
state majorize those of the input state.

Thus, applying the majorization condition to the states (\ref{eqn7})
and the initial TMSS yields
\begin{equation}
\left (\lambda e^{2 \kappa_T \mbox{Img}(O_W) } \right)^{\ell} >
\lambda^{\ell}, \label{eqn14}
\end{equation}
$\forall \ell \in [1, \infty]$. The only way to satisfy
(\ref{eqn14}) is if the imaginary part of $O_W$ is positive for all
$\ell$ (assuming $\kappa_T>0$). Entanglement concentration
then only occurs if
\begin{equation}
\mbox{Img} \left ( \frac{\langle \Phi_2 \vert \hat O_C \vert \Phi_1
\rangle}{\langle \Phi_2 \vert \Phi_1 \rangle} \right ) > 0.
\label{eqn15}
\end{equation}
Consequently, the protocol provides a success condition
for entanglement concentration based on the properties of the weak
value of a particular observable. This in conjugation with condition
(\ref{eqn9}) provides a number of constraints that the interaction
Hamiltonian ${\hat H}_I$, the pre-selected and post-selected ancilla
states $\vert \Phi_1 \rangle$ and $\vert \Phi_2 \rangle$ and the
observable $\hat O_C$ must obey in order to produce entanglement
concentration of the TMSS. It is interesting to note that the weak
condition (\ref{eqn9}) coupled with the requirements of the {\it
Procrustean} method are all that is required to preserve the
gaussian character of the TMSS.

\section{Examples}

We now demonstrate that previously discovered protocols of this type
can emerge as special examples of the general model advocated here.
We will also calculate the associate weak values and demonstrate
that the weakness condition is satisfied. The previous schemes
\cite{fiurasek:2003} and \cite{menzies:2006}, required that Bob's
half of the TMSS be mixed with an ancillary coherent state $\vert
\alpha \rangle$, where $\alpha \in \Re$ and $\alpha > 0$, in a
non-linear medium exhibiting the cross Kerr effect $\hat H_I = \hbar
\kappa(t) \hat n_B \hat n_C$ before being subjected to a measurement
and post-selection condition. Using the success condition
(\ref{eqn15}), we can derive a constraint on the possible
post-selected ancilla states which will allow us to select
measurement strategies the lead to Gaussian-preserving entanglement
concentration. Thus, we are interested in the weak values of the
number operator $\hat n_C$:
\begin{equation}
n_W = \frac{ \langle \Phi_2 \vert \hat n_C \vert \alpha
\rangle}{\langle \Phi_2 \vert \alpha \rangle} =
\frac{e^{-\alpha^2/2} \alpha \partial_{\alpha} \left (
e^{\alpha^2/2} \langle \Phi_2 \vert \alpha \rangle \right )}{\langle
\Phi_2 \vert \alpha \rangle}. \label{eqn16}
\end{equation}
The second equality in (\ref{eqn16}) follows from $\alpha
\partial_{\alpha} (\alpha^n) = n \alpha^n$. Furthermore, if we
assume
\begin{equation}
\langle \Phi_2 \vert \alpha \rangle = R(\alpha) e^{i
\theta(\alpha)}, \label{eqn17}
\end{equation}
where $R(\alpha)$ and $\theta(\alpha)$ are the magnitude and phase
of the scalar product of $\langle \Phi_2 \vert \alpha \rangle$, then
after some algebra (\ref{eqn16}) can be written as
\begin{equation}
n_W = \alpha^2 + \frac{\alpha}{R(\alpha)} \frac{\partial R}{\partial
\alpha} + i \alpha \frac{\partial \theta}{\partial \alpha}.
\label{eqn18}
\end{equation}
Consequently, the success condition requires that $\mbox{Img}(n_W) >
0 \Leftrightarrow \alpha \partial_{\alpha} \theta(\alpha) > 0$.
Thus, the only variants of this family of protocols which achieve
the desired effect are those where the phase of $\langle \Phi_2
\vert \alpha \rangle$ is a monotonic increasing function of
$\alpha$. This prediction allows us to recover previously suggested
protocols and uncover new variants.

{\it Fiur$\grave{a}$$\check{s}$ek, Mi$\check{s}$ta and Filip} (2003)
\cite{fiurasek:2003}. In this scheme, the ancillary coherent state
is projected onto $\vert \beta \rangle = \vert \vert \beta \vert
e^{i \phi} \rangle$ via eight-port-Homodyne detection. This example
prevails due to the over-complete nature of coherent states
\begin{equation}
\langle \beta \vert \alpha \rangle = e^{-\alpha^2/2} e^{-\vert \beta
\vert^2/2} e^{\alpha \beta^*}, \label{eqn19}
\end{equation}
where it is clear that the phase of the above is a
monotonic increasing function of $\alpha$ only if the imaginary part
of $\beta$ is negative. This also follows from
\begin{equation}
\mbox{Img}(n_W) = \alpha \partial_{\alpha} \theta(\alpha) = - \alpha
\vert \beta \vert \sin \phi. \label{eqn20}
\end{equation}
Hence, the success condition for this protocol is given by
$\pi < \phi < 2 \pi$ and only states post-selected with respect to
this condition will allow the desired effect. The weakness condition
is then given as
\begin{equation}
\lambda^n \left ( \frac{\langle \beta \vert e^{- i \kappa_T n \hat
n_C } \vert \alpha \rangle}{\langle \beta \vert \alpha \rangle} -
e^{- i \kappa_T  n \alpha \beta^*} \right ) = 0 \; \; \forall n \in
[0,\infty), \label{eqn21}
\end{equation}
Using the identity  $\exp \left ( \sigma \hat a^{\dagger}
\hat a \right ) = : \exp \left ( \{ e^{\sigma} - 1 \} \hat
a^{\dagger} \hat a \right ) :$ \cite{barnett&radmore:1997}, then
(\ref{eqn21}) becomes
\begin{equation}
\lambda^n \left ( e^{(e^{-i \kappa_T n} - 1) \beta^* \alpha} - e^{-i
\kappa_T n \beta^* \alpha } \right ) = 0 \; \; \forall n \in
[0,\infty). \label{eqn22}
\end{equation}
The above is true if $\kappa_T <<1$ such that $e^{- i
\kappa_t n} \approx 1 - i \kappa_T n$, which only holds for
sufficiently small $n$. Thus, for small values of $n$, (\ref{eqn22})
is satisfied. However, for large values of $n$ where $e^{- i
\kappa_t n} \neq 1 - i \kappa_T n$, (\ref{eqn22}) still holds
because $\lambda < 1$ and hence $\lambda^n \rightarrow 0$ for
progressively larger $n$. Thus, the weakness condition requires a
balancing act between the non-linear coupling and the squeezing of
the input TMSS. The authors of
\cite{fiurasek:2003} arrive at the same conclusion.

{\it Menzies and Korolkova} (2006) \cite{menzies:2006}. Here
balanced Homodyne detection is employed by Bob, in other words, the
post-selected state of the ancilla is the quadrature eigenstate
$\vert x_{\phi} \rangle = \vert \Phi_2 \rangle$ where
$\hat{X}_{\phi} \vert x_{\phi} \rangle = x_{\phi} \vert x_{\phi}
\rangle$ and $\hat{X}_{\phi} = 2^{-1/2} ( e^{i \phi} \hat
a^{\dagger} + e^{-i \phi} \hat a )$. Once again, this protocol works
because of the nature of the overlap between the pre- and
post-selected states. In this case, we have
\cite{barnett&radmore:1997}
\begin{equation}
\langle x_{\phi} \vert \alpha \rangle = \pi^{-1/4} e^{- x_{\phi}^2/2
+ \sqrt{2} e^{ - i \phi }x_{\phi} \alpha - e^{- 2 i \phi}
\alpha^2/2}, \label{eqn23}
\end{equation}
then the imaginary part of the weak value is
\begin{equation}
\mbox{Img}(n_W) =\alpha \partial_{\alpha} \theta = \sqrt{2} \alpha
\sin \phi x_{\phi} - \alpha^2 \sin(2 \phi), \label{eqn24}
\end{equation}
and so condition (\ref{eqn15}) translates to
\begin{equation}
\mbox{Img}(n_W) > 0 \Leftrightarrow x_{\phi} > \sqrt{2} \alpha \cos
\phi. \label{eqn25}
\end{equation}
Every possible quadrature measurement has its own success
condition for entanglement concentration. This condition defines the
post-selection criterion. The weakness condition for this protocol
is given as
\begin{equation}
\lambda^n \left ( \frac{\langle x_{\phi} \vert e^{-i \kappa_T n \hat
n_C } \vert \alpha \rangle}{\langle x_{\phi} \vert \alpha \rangle} -
e^{- i \kappa_T n n_W} \right ) =0, \forall n \in [0, \infty).
\label{eqn26}
\end{equation}
This can be re-expressed as ($\forall \: n \in [0, \infty)$):
\begin{eqnarray}
\lambda^n \left [ \exp{ \left ( \sqrt{2} x_{\phi} \alpha e^{i \phi -
i \kappa_T n} - \alpha^2 \frac{e^{2 i \phi - 2 i \kappa_T n}}{2}
\right) } \right . - \nonumber \\ \left .\exp{ \left (- i \kappa_T n
\left \{ \sqrt{2} x_{\phi} \alpha e^{i \phi} - \alpha^2 e^{2 i \phi}
\right \} \right) } \right ] =0. \nonumber
\end{eqnarray}
So, just as for the previous example, we see that (\ref{eqn26}) is
equivalent to (\ref{eqn22}). Thus, both schemes require the
balancing between the initial {\it Schmidt} coefficients and the
magnitude of the non-linear coupling.

{\it Squeezed vacuum post-selection scheme}. To generate further
examples, we simply need to identify further quantum optical states
that satisfy $\partial \alpha \theta(\alpha) > 0$. An immediate and
obvious choice is given by selecting the post-selected state as a
single mode squeezed vacuum $\vert \Phi_2 \rangle = \vert r e^{i
\phi} \rangle$ since \cite{barnett&radmore:1997}
\begin{equation}
\langle r e^{i \phi} \vert \alpha \rangle = \sqrt{\mbox{sech} r}
\exp \left ( - \frac{\alpha^2}{2} \{ 1 + e^{- i \phi} \tanh r\}
\right ). \label{eqn27}
\end{equation}
Consequently, the phase of the above overlap is given as
\begin{equation}
\theta(\alpha) = \frac{\alpha^2}{2} \sin \phi \tanh r, \label{eqn28}
\end{equation}
and hence, in this example, the success condition
(\ref{eqn15}) is
\begin{equation}
\mbox{Img}(n_W) =  \alpha^2 \tanh r \sin \phi > 0 \Leftrightarrow 0
< \phi < \pi/2. \label{eqn29}
\end{equation}
The weak condition is expressed as
\begin{equation}
\lambda^n \left ( \frac{\langle r e^{i \phi} \vert e^{-i \kappa_T n
\hat n_C } \vert \alpha \rangle}{\langle r e^{i \phi} \vert \alpha
\rangle} - e^{- i \kappa_T n n_W} \right ) =0, \; \; \forall n \in
[0, \infty), \label{eqn30}
\end{equation}
where the first term on the LHS is $\exp \left
(-\frac{\alpha^2}{2} \{ e^{ 2 i n \kappa_T} - 1 \}e^{- i \phi} \tanh
r \right )$ and the second is $\exp \left ( - i \kappa_T n \alpha^2
e^{- i \phi} \tanh r \right )$. Clearly (\ref{eqn30}) can only be
satisfied if $\kappa_T n << 1$. Note that for large $n$,
(\ref{eqn30}) holds because $\lambda^n \rightarrow 1$. In this
example, the power of the weak value approach is evident due to its
ability to provide an elegant shortcut to the required operation
conditions.

\section{Concluding Remarks}

In conclusion, we have illustrated an application of weak
measurements for a Gaussian preserving entanglement concentration.
In particular, we have provided a general weak measurement model
that allows for the probabilistic transformation of an input TMSS
$\vert \zeta(\lambda) \rangle$ to an output $\vert \zeta(\lambda')
\rangle$. This model allows several useful features. Firstly, we can
formulate an entanglement concentration success condition dependant
on the magnitude of the imaginary part of the weak value. Moreover,
this success condition also puts a constrain on the initial ancilla
states and the subsequent measurement strategy employed. Secondly,
the weakness criterion in conjunction with the requirements of the
{\it Procrustean} method guarantee that the entanglement
concentration will be Gaussian preserving. As a consequence of the
universality of the model provided here, it is a simple matter to
determine novel examples of this protocol by appealing to the
imaginary part of the ancilla's weak value. Indeed, we attempt to
justify this viewpoint by providing another example of this
entanglement concentration protocol inspired purely from
observations of the imaginary weak value.

Despite the advances offered by adopting the weak measurement
paradigm here, there remain a number of outstanding problems. For
example, can we generalize this model to account for type-preserving
{\it Procrustean} entanglement concentration for arbitrary
continuous-variable pure bipartite entangled states? Furthermore, we
note that the weak condition itself is never totally satisfied.
Indeed, whereas it is easy to satisfy (\ref{eqn9}) for small and
large $n$ by requiring that $\lambda < 1$ and $\kappa_T << 1$, it
is, however, not clear if weakness condition is true for
intermediate values of $n$. It is then natural to enquire the
consequences of, if any, such violations to the ability of the
protocol to deliver Gaussian preserving entanglement concentration.
In principle, we would like to obtain a quantitative relation
between the magnitude of the violation and the ability of
(\ref{eqn15}) to act as a success condition. The weakness condition
allows the Gaussian preservation and so failure of the condition
results in a non-Gaussian output state. Thus, we could measure the
violation of (\ref{eqn9}) by measuring the extent of the
non-Gaussian character of the output state. However, obtaining such
a quantitative relation is highly non-trivial as the features of the
non-Gaussian entangled states cannot, in many cases, be calculated
in an analytical fashion. Consequently, the formulation of such a
relation remains a goal for future research.

{\it Acknowledgements}.
This work was supported by the {\it Engineering and Physical
Sciences Research Council} and by the EU project FP6-511004
COVAQIAL.

\bibliographystyle{unsrt}
\small \small \small \small
\bibliography{weak}

\end{document}